\documentclass[prd,preprintnumbers,twocolumn,amsmath,nofootinbib,amssymb]{revtex4}
\usepackage{graphicx,color,dcolumn,booktabs,bm}
\usepackage{longtable,lscape}
\usepackage{txfonts}
\usepackage{overpic}
\usepackage{amssymb}
\usepackage{epstopdf}
\usepackage{indentfirst}
\usepackage{feynmf}   
\usepackage{slashed}  
\usepackage{cases}
\usepackage{color}
\usepackage{float}
\usepackage{multirow}
\usepackage{ulem}
\usepackage{graphicx,color,dcolumn,booktabs,bm}
\usepackage{epsfig,dsfont,amssymb,amsmath,amsfonts,amsbsy,mathrsfs}

\graphicspath{{Figures/}} %

\usepackage{hyperref}
\hypersetup{colorlinks,citecolor=blue,anchorcolor=red,menucolor=red, linkcolor=red,filecolor=red,runcolor=red,urlcolor=blue,frenchlinks=red}

\makeatletter
\@addtoreset{equation}{section}
\makeatother

\allowdisplaybreaks

\begin{document}

\title{Prediction of hidden-charm pentaquarks with double strangeness}

\author{Fu-Lai Wang$^{1,2}$}
\email{wangfl2016@lzu.edu.cn}
\author{Rui Chen$^{3}$}
\author{Xiang Liu$^{1,2,4}$\footnote{Corresponding author}}
\email{xiangliu@lzu.edu.cn}
\affiliation{$^1$School of Physical Science and Technology, Lanzhou University, Lanzhou 730000, China\\
$^2$Research Center for Hadron and CSR Physics, Lanzhou University and Institute of Modern Physics of CAS, Lanzhou 730000, China\\
$^3$Center of High Energy Physics, Peking University, Beijing
100871, China\\
$^4$Lanzhou Center for Theoretical Physics, Lanzhou University, Lanzhou 730000, China}

\begin{abstract}
Inspired by the recent evidence of $P_{cs}(4459)$ reported by LHCb, we continue to perform the investigation of hidden-charm molecular pentaquarks with double strangeness, which are composed of an $S$-wave charmed baryon $\Xi_c^{(\prime,*)}$ and an $S$-wave anti-charmed-strange meson $\bar{D}_s^{(*)}$. Both the $S$-$D$ wave mixing effect and the coupled channel effect are taken into account in realistic calculation. A dynamics calculation shows that there may exist two types of hidden-charm molecular pentaquark with double strangeness, i.e., the $\Xi_{c}^{*}\bar D_s^*$ molecular state with $J^P={5}/{2}^{-}$ and the $\Xi_{c}^{\prime}\bar D_s^*$ molecular state with $J^P={3}/{2}^{-}$. According to this result, we strongly suggest the experimental exploration of hidden-charm molecular pentaquarks with double strangeness. Facing such opportunity, obviously the LHCb will have great potential to hunt for them, with the data accumulation at Run III and after High-Luminosity-LHC upgrade.
\end{abstract}

\maketitle

\section{Introduction}\label{sec1}
Very recently, the LHCb Collaboration reported the evidence of a possible hidden-charm pentaquark with strangeness $P_{cs}(4459)$ by analyzing the $\Xi_b^-\to J/\psi K^- \Lambda$ process, where this enhancement structure existing in the $J/\psi\Lambda$ invariant mass spectrum has the mass $4458.8\pm2.9^{+4.7}_{-1.1}$ MeV and the width $17.3\pm6.5^{+8.0}_{-5.7}$ MeV \cite{lhcb}. In fact, before this announcement, several theoretical groups \cite{Chen:2016ryt,Wu:2010vk,Hofmann:2005sw,Anisovich:2015zqa,Wang:2015wsa,Feijoo:2015kts,Lu:2016roh,Xiao:2019gjd,Shen:2020gpw,Chen:2015sxa,Zhang:2020cdi,Wang:2019nvm} predicted the existence of hidden-charm pentaquarks with strangeness and suggested experimentalist to carry out the search for them. After reporting $P_{cs}(4459)$, there were also theoretical studies to decode $P_{cs}(4459)$, where $P_{cs}(4459)$ can be assigned as the $\Xi_c \bar{D}^*$ molecular state \cite{Chen:2020uif,Peng:2020hql,1830432,1830426,1830449}.

In fact, the above experimental investigation of pentaquark is a continuation of the past. In 2015, the LHCb Collaboration released the observation of $P_c(4380)$ and $P_c(4450)$ in the $J/\psi p$ invariant mass spectrum of the $\Lambda_b \to J/\psi p K$ process \cite{Aaij:2015tga}. In 2019, based on more collected data, the LHCb Collaboration again analyzed the same process and found a new $P_c(4312)$ and indicated that $P_c(4450)$ contains two substructures $P_c(4440)$ and $P_{c}(4457)$ \cite{Aaij:2019vzc}, which may provide strong evidence of hidden-charm molecular pentaquarks existing in nature \cite{Li:2014gra,Karliner:2015ina,Wu:2010jy,Wang:2011rga,Yang:2011wz,Wu:2012md,Chen:2015loa}.

\begin{figure}[!htbp]
\includegraphics[width=8.3cm,keepaspectratio]{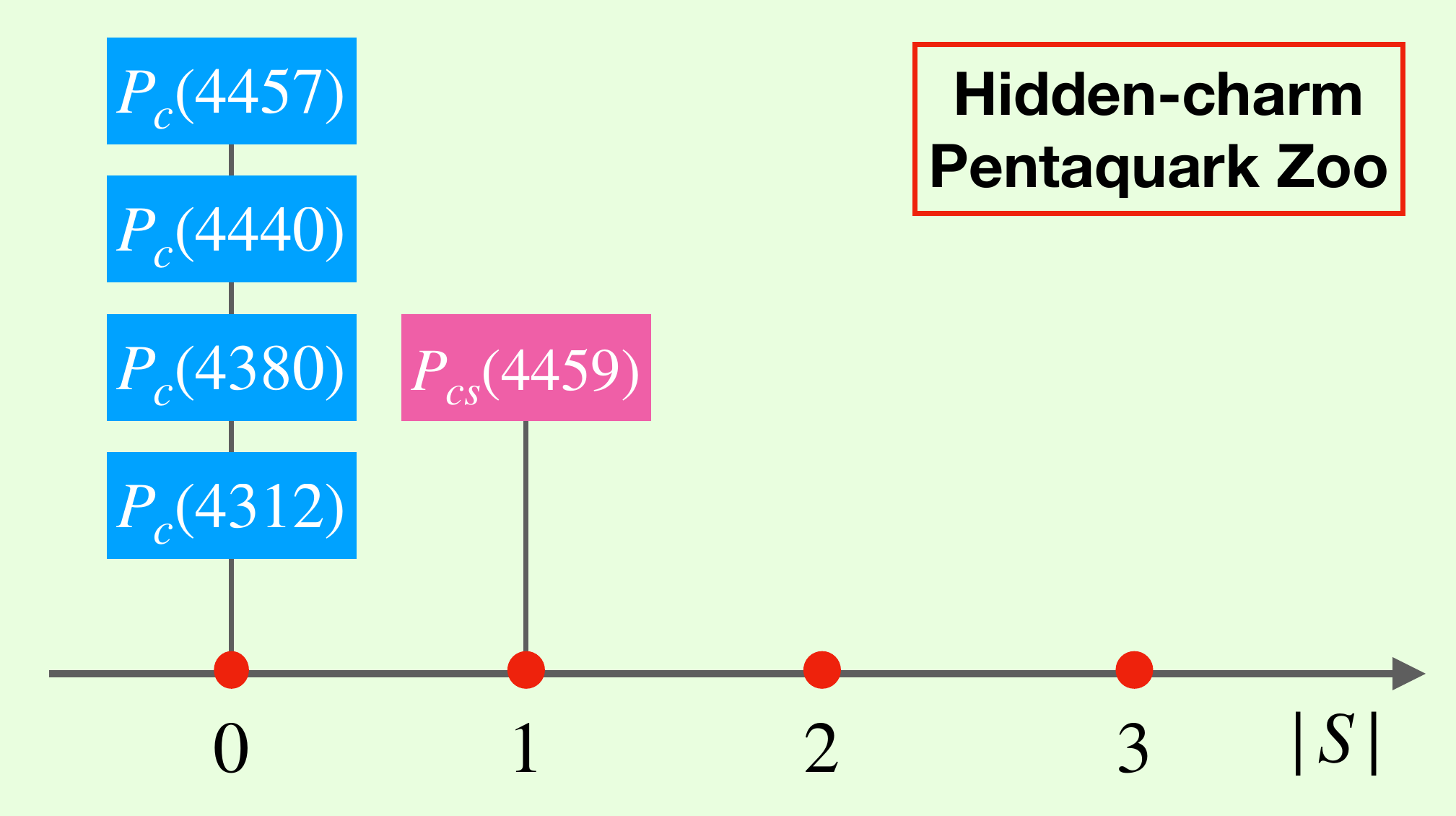}
\caption{The hidden-charm pentaquark zoo with different values of strangeness.}
\label{fig1}
\end{figure}

When facing such new progress on exploring hidden-charm pentaquarks \cite{Chen:2016qju,Liu:2019zoy,Olsen:2017bmm,Guo:2017jvc}, we naturally expect that there should form a zoo of hidden-charm pentaquarks which is composed of different kinds of hidden-charm pentaquarks (see Fig. \ref{fig1}). If considering the difference of strangeness in hidden-charm pentaquarks, we want to further get the information of hidden-charm pentaquarks with double strangeness $|S|=2$ when $P_{cs}(4459)$ with strangeness $|S|=1$ was announced, which will be the main task of this work.

In this work, we study hadronic molecular states composed of an $S$-wave charmed baryon $\Xi_c^{(\prime,*)}$ and an $S$-wave anti-charmed-strange meson $\bar{D}_s^{(*)}$. Here, $\Xi_c$ with $J^P=1/2^+$ denotes $S$-wave charmed baryon in $\bar{3}_F$ flavor representation, while $S$-wave charmed baryons $\Xi_c^\prime$ with $J^P=1/2^+$ and $\Xi_c^*$ with $J^P=3/2^+$ are in $6_F$ flavor representation. For obtaining the effective potentials of these discussed systems, we adopt the one-boson-exchange (OBE) model. With the extracted effective potentials, we may find the bound state solutions of the discussed systems, and further conclude whether there exist possible hidden-charm molecular pentaquarks with double strangeness. By this effort, we may give theoretical suggestion of performing the search for them in experiment.

For Large Hadron Collider (LHC), Run III will be carried out in near future. After that, the High-Luminosity-LHC upgrade will eventually collect an integrated luminosity of 300 $\rm {fb^{-1}}$ of data in pp collisions at a centre-of-mass energy of 14 TeV \cite{Bediaga:2018lhg}. Thus, we have reason to believe that exploring these suggested hidden-charm molecular pentaquarks with double strangeness will be an interesting research issue full of opportunity, especially at the LHCb.

This paper is organized as the follows. After introduction, the detailed calculation of the interactions of these discussed $\Xi_c^{(\prime,*)}\bar{D}_s^{(*)}$ systems will be given in Sec. \ref{sec2}. With this preparation, we present the bound state properties for these possible hidden-charm pentaquark molecules with double strangeness in Sec. \ref{sec3}. Finally, this work ends with a short summary in Sec. \ref{sec4}.

\section{Effective potentials of the $\Xi_c^{(\prime,*)}\bar{D}_s^{(*)}$ systems}\label{sec2}
In this work, we mainly study the interactions between an $S$-wave charmed baryon $\Xi_c^{(\prime,*)}$ and an $S$-wave anti-charmed-strange meson $\bar{D}_s^{(*)}$ in the OBE model, we include the contribution from the pseudoscalar meson, scalar meson, and vector meson exchanges in our concrete calculation, which is often adopted to study the hadron interactions \cite{Chen:2016qju,Liu:2019zoy}.

\subsection{Deducing the effective potentials}
As shown in Refs. \cite{Wang:2020dya,Wang:2019nwt}, there usually exist three typical steps for calculating the effective potentials of these discussed $\Xi_c^{(\prime,*)}\bar{D}_s^{(*)}$ systems, we will give a brief introduction in the following.

Firstly, we can obtain the scattering amplitudes $\mathcal{M}(h_1h_2\to h_3h_4)$ for the $h_1h_2\to h_3h_4$ processes with the effective Lagrangian approach. Here, we need to construct the relevant effective Lagrangians in the present work.

If considering the heavy quark symmetry, chiral symmetry \cite{Wise:1992hn,Casalbuoni:1992gi,Casalbuoni:1996pg,Yan:1992gz}, and hidden local symmetry \cite{Bando:1987br,Harada:2003jx}, the effective Lagrangians for depicting the heavy hadrons $\bar{D}_s^{(*)}/\mathcal{B}_c^{(*)}$ coupling with the light scalar, pesudoscalar, and vector mesons read as \cite{Ding:2008gr,Chen:2017xat}
\begin{eqnarray}
\mathcal{L}_{H}&=&g_S\langle \bar{H}^{(\overline{Q})}_a\sigma H^{(\overline{Q})}_b\rangle+ig\langle \bar{H}^{(\overline{Q})}_a\gamma_{\mu}{\mathcal A}_{ab}^{\mu}\gamma_5H^{(\overline{Q})}_b\rangle\nonumber\\
  &&-i\beta\langle \bar{H}^{(\overline{Q})}_av_{\mu}\left(\mathcal{V}^{\mu}-\rho^{\mu}\right)_{ab}H^{(\overline{Q})}_b\rangle\nonumber\\
  &&+i\lambda\langle \bar{H}^{(\overline{Q})}_a\sigma_{\mu\nu}F^{\mu\nu}(\rho)H^{(\overline{Q})}_b\rangle,\nonumber\\
\mathcal{L}_{\mathcal{B}_{\bar{3}}} &=& l_B\langle\bar{\mathcal{B}}_{\bar{3}}\sigma\mathcal{B}_{\bar{3}}\rangle +i\beta_B\langle\bar{\mathcal{B}}_{\bar{3}}v^{\mu}(\mathcal{V}_{\mu}-\rho_{\mu})\mathcal{B}_{\bar{3}}\rangle,\nonumber\\
\mathcal{L}_{\mathcal{B}^{(*)}_6} &=&  l_S\langle\bar{\mathcal{S}}_{\mu}\sigma\mathcal{S}^{\mu}\rangle
         -\frac{3}{2}g_1\varepsilon^{\mu\nu\lambda\kappa}v_{\kappa}\langle\bar{\mathcal{S}}_{\mu}{\mathcal A}_{\nu}\mathcal{S}_{\lambda}\rangle\nonumber\\
  &&+i\beta_{S}\langle\bar{\mathcal{S}}_{\mu}v_{\alpha}\left(\mathcal{V}^{\alpha}-\rho^{\alpha}\right) \mathcal{S}^{\mu}\rangle
         +\lambda_S\langle\bar{\mathcal{S}}_{\mu}F^{\mu\nu}(\rho)\mathcal{S}_{\nu}\rangle,\nonumber\\
\mathcal{L}_{\mathcal{B}_{\bar{3}}\mathcal{B}^{(*)}_6} &=& ig_4\langle\bar{\mathcal{S}^{\mu}}{\mathcal A}_{\mu}\mathcal{B}_{\bar{3}}\rangle
         +i\lambda_I\varepsilon^{\mu\nu\lambda\kappa}v_{\mu}\langle \bar{\mathcal{S}}_{\nu}F_{\lambda\kappa}\mathcal{B}_{\bar{3}}\rangle+h.c..
\end{eqnarray}
Here, $v=(1,\bm{0})$ is the four velocity under the non-relativistic approximation. The super-field $H^{(\overline{Q})}_a$ is given by $H^{(\overline{Q})}_a=\left(\bar{D}^{*(\overline{Q})\mu}_{a}\gamma_{\mu}-\bar{D}^{(\overline{Q})}_a\gamma_5\right)\frac{1-\slash \!\!\!v}{2}$, which is expressed as a combination of the vector anti-charmed-strange meson $\bar{D}^{*}_s$ with $J^P=1^-$ and the pseudoscalar anti-charmed-strange meson $\bar{D}_s$ with $J^P=0^-$. And the super-field $\mathcal{S}_{\mu}$ can be written as $\mathcal{S}_{\mu} =-\sqrt{\frac{1}{3}}(\gamma_{\mu}+v_{\mu})\gamma^5\mathcal{B}_6+\mathcal{B}_{6\mu}^*$, which includes the charmed baryons $\mathcal{B}_6$ with $J^P=1/2^+$ and $\mathcal{B}^*_6$ with $J^P=3/2^+$ in the $6_F$ flavor representation. Matrices $\mathcal{B}_{\bar{3}}$ and $\mathcal{B}_6^{(*)}$ are given by
\begin{eqnarray}
\mathcal{B}_{\bar{3}} = \left(\begin{array}{ccc}
        0    &\Lambda_c^+      &\Xi_c^+\\
        -\Lambda_c^+       &0      &\Xi_c^0\\
        -\Xi_c^+      &-\Xi_c^0     &0
\end{array}\right),
\mathcal{B}_6^{(*)} = \left(\begin{array}{ccc}
         \Sigma_c^{{(*)}++}                  &\frac{\Sigma_c^{{(*)}+}}{\sqrt{2}}     &\frac{\Xi_c^{(',*)+}}{\sqrt{2}}\\
         \frac{\Sigma_c^{{(*)}+}}{\sqrt{2}}      &\Sigma_c^{{(*)}0}    &\frac{\Xi_c^{(',*)0}}{\sqrt{2}}\\
         \frac{\Xi_c^{(',*)+}}{\sqrt{2}}    &\frac{\Xi_c^{(',*)0}}{\sqrt{2}}      &\Omega_c^{(*)0}
\end{array}\right),
\end{eqnarray}
respectively.  In addition, the axial current $\mathcal{A}_\mu$, the vector current ${\cal V}_{\mu}$, the vector meson field $\rho_{\mu}$, and the vector meson strength tensor $F_{\mu\nu}(\rho)$ are defined as
\begin{eqnarray}
{\mathcal A}_{\mu}&=&\frac{1}{2}\left(\xi^{\dagger}\partial_{\mu}\xi-\xi\partial_{\mu}\xi^{\dagger}\right)_{\mu},\nonumber\\
{\mathcal V}_{\mu}&=&\frac{1}{2}\left(\xi^{\dagger}\partial_{\mu}\xi+\xi\partial_{\mu}\xi^{\dagger}\right)_{\mu},\nonumber\\
\rho_{\mu}&=&\frac{i{g_V}}{{\sqrt{2}}}\mathbb{V}_{\mu},\nonumber\\
F_{\mu\nu}(\rho)&=&\partial_{\mu}\rho_{\nu}-\partial_{\nu}\rho_{\mu}+\left[\rho_{\mu},\rho_{\nu}\right],
\end{eqnarray}
respectively. Here, $\xi=\exp(i\mathbb{P}/f_\pi)$ with $f_\pi$ as the pion decay constant, ${\mathbb{P}}$ and $\mathbb{V}_{\mu}$ correspond to the light pseudoscalar and vector meson matrices, which are expressed as
\begin{eqnarray}
\left.\begin{array}{c}
{\mathbb{P}} = {\left(\begin{array}{ccc}
       \frac{\pi^0}{\sqrt{2}}+\frac{\eta}{\sqrt{6}} &\pi^+ &K^+\\
       \pi^-       &-\frac{\pi^0}{\sqrt{2}}+\frac{\eta}{\sqrt{6}} &K^0\\
       K^-         &\bar K^0   &-\sqrt{\frac{2}{3}} \eta     \end{array}\right)},\\
{\mathbb{V}}_{\mu} = {\left(\begin{array}{ccc}
       \frac{\rho^0}{\sqrt{2}}+\frac{\omega}{\sqrt{2}} &\rho^+ &K^{*+}\\
       \rho^-       &-\frac{\rho^0}{\sqrt{2}}+\frac{\omega}{\sqrt{2}} &K^{*0}\\
       K^{*-}         &\bar K^{*0}   & \phi     \end{array}\right)}_{\mu},
\end{array}\right.
\end{eqnarray}
respectively. By expanding the compact effective Lagrangians to the leading order of the pseudo-Goldstone field, the detailed effective Lagrangians for the heavy hadrons $\bar{D}_s^{(*)}/\mathcal{B}_c^{(*)}$ and the exchanged light mesons can be obtained (see Refs. \cite{Chen:2017xat,Chen:2018pzd,Wang:2020dya} for more details).

Secondly, the effective potentials in the momentum space can be related to the scattering amplitudes via the Breit approximation \cite{Breit:1929zz,Breit:1930zza} and the non-relativistic normalization, where the general relation can be explicitly expressed as
\begin{eqnarray}
\mathcal{V}^{h_1h_2\to h_3h_4}_E(\bm{q})=-\frac{\mathcal{M}(h_1h_2\to h_3h_4)}{\sqrt{\prod_i 2 m_i}}.
\end{eqnarray}
In the above relation, $\mathcal{V}^{h_1h_2\to h_3h_4}_E(\bm{q})$ is the effective potential in the momentum space, and $m_{i}\,(i=h_1,\,h_2,\,h_3,\,h_4)$ represents the mass of the initial and final states.

Thirdly, we discuss the bound state properties of the $\Xi_c^{(\prime,*)}\bar{D}_s^{(*)}$ systems by solving the coupled channel Schr$\ddot{\rm o}$dinger equation in the coordinate space. Thus, the effective potentials in the coordinate space $\mathcal{V}^{h_1h_2\to h_3h_4}_E(\bm{r})$ can be obtained by performing Fourier transformation, i.e.,
\begin{eqnarray}
\mathcal{V}^{h_1h_2\to h_3h_4}_E(\bm{r}) =\int \frac{d^3\bm{q}}{(2\pi)^3}e^{i\bm{q}\cdot\bm{r}}\mathcal{V}^{h_1h_2\to h_3h_4}_E(\bm{q})\mathcal{F}^2(q^2,m_E^2).
\end{eqnarray}
As a general rule, the form factor should be introduced in the interaction vertex to describe the off-shell effect of the exchanged light mesons and reflect the inner structure effect of the discussed hadrons. In this work, we introduce the monopole type form factor $\mathcal{F}(q^2,m_E^2) = (\Lambda^2-m_E^2)/(\Lambda^2-q^2)$ \cite{Tornqvist:1993ng,Tornqvist:1993vu}. Here, $\Lambda$, $m_E$, and $q$ are the cutoff parameter, the mass, and the four momentum of the exchanged light mesons, respectively. In our calculation, we attempt to find bound state solutions by varying the cutoff parameter in the range of $1.00$-$5.00~{\rm GeV}$ \cite{Chen:2018pzd,Wang:2019aoc}. In particular, we need to emphasize that the loosely bound state with the cutoff value around 1.00 GeV may be the most promising molecular candidate, which is widely accepted as a reasonable input parameter from the experience of studying the deuteron \cite{Tornqvist:1993ng,Tornqvist:1993vu,Wang:2019nwt,Chen:2017jjn}. Of course, when judging whether the loosely bound state is an ideal hadronic molecular candidate, we also need to specify that the reasonable binding energy should be at most tens of MeV, and the typical size should be larger than the size of all the included component hadrons \cite{Chen:2017xat}. 

\subsection{Constructing the spin-orbital and flavor wave functions}
In order to obtain the effective potentials of these investigated systems, we need to construct the spin-orbital and flavor wave functions. For these discussed $\Xi_c^{(\prime,*)}\bar{D}_s^{(*)}$ systems, their spin-orbital wave functions can be expressed as
\begin{eqnarray}
|\mathcal{B}_{c}\bar{D}_{s}({}^{2S+1}L_{J})\rangle&=&\sum_{m,m_L}C^{J,M}_{\frac{1}{2}m,Lm_L}\chi_{\frac{1}{2}m}|Y_{L,m_L}\rangle,\nonumber\\
|\mathcal{B}_{c}^*\bar{D}_{s}({}^{2S+1}L_{J})\rangle&=&\sum_{m,m_L}C^{J,M}_{\frac{3}{2}m,Lm_L}\Phi_{\frac{3}{2}m}|Y_{L,m_L}\rangle,\nonumber\\
|\mathcal{B}_{c}\bar{D}_{s}^*({}^{2S+1}L_{J})\rangle&=&\sum_{m,m',m_S,m_L}C^{S,m_S}_{\frac{1}{2}m,1m'}C^{J,M}_{Sm_S,Lm_L}\chi_{\frac{1}{2}m}\epsilon_{m'}^{\mu}|Y_{L,m_L}\rangle,\nonumber\\
|\mathcal{B}_{c}^{*}\bar{D}_{s}^*({}^{2S+1}L_{J})\rangle&=&\sum_{m,m',m_S,m_L}C^{S,m_S}_{\frac{3}{2}m,1m'}C^{J,M}_{Sm_S,Lm_L}\Phi_{\frac{3}{2}m}\epsilon_{m'}^{\mu}|Y_{L,m_L}\rangle.\nonumber\\
\end{eqnarray}
In the above expressions, the notations $\mathcal{B}_{c}$ and $\mathcal{B}_{c}^*$ denote the charmed baryons $\Xi_c^{(\prime)}$ and $\Xi_c^{*}$, respectively. The constant $C^{e,f}_{ab,cd}$ is the Clebsch-Gordan coefficient, and $|Y_{L,m_L}\rangle$ stands for the spherical harmonics function. The polarization vector $\epsilon_{m}^{\mu}\,(m=0,\,\pm1)$ with spin-1 field is written as $\epsilon_{\pm}^{\mu}= \left(0,\,\pm1,\,i,\,0\right)/\sqrt{2}$ and $\epsilon_{0}^{\mu}= \left(0,0,0,-1\right)$ in the static limit. The $\chi_{\frac{1}{2}m}$ denotes the spin wave function for the charmed baryons $\Xi_c^{(\prime)}$ with spin $S={1}/{2}$, and the polarization tensor $\Phi_{\frac{3}{2}m}$ for the charmed baryon $\Xi_c^{*}$ with spin $S={3}/{2}$ has the form of $\Phi_{\frac{3}{2}m}=\sum_{m_1,m_2}C^{{3}/{2},m}_{{1}/{2}m_1;1m_2}\chi_{\frac{1}{2},m_1}\epsilon_{m_2}^{\mu}$.

And then, the flavor wave functions $|I,I_3\rangle$ of the $\Xi_c^{(\prime,*)}\bar{D}_s^{(*)}$ systems have the form of $|1/2,1/2\rangle=|\Xi_c^{(\prime,*)+}{D}_s^{(*)-}\rangle$ and $|1/2,-1/2\rangle=|\Xi_c^{(\prime,*)0}{D}_s^{(*)-}\rangle$, where $I$ and $I_3$ are isospin and its third component of the investigated systems.

Specifically, the study of the deuteron indicates that the $S$-$D$ wave mixing effect may play an important role in the formation of the loosely bound states \cite{Wang:2019nwt}. Thus, the $S$-$D$ wave mixing effect is also considered in this work, and the possible channels involved in our calculation are summarized in Table~\ref{spin-orbit wave functions}.
\renewcommand\tabcolsep{0.13cm}
\renewcommand{\arraystretch}{1.50}
\begin{table}[!htpb]
\centering
\caption{The possible channels involved in our calculation. Here, ``$...$" means that the $S$-wave components for the corresponding channels do not exist.}\label{spin-orbit wave functions}
\begin{tabular}{c|cccc}\toprule[1pt]\toprule[1pt]
 $J^{P}$&$\Xi_c^{(\prime)}\bar{D}_s$&$\Xi_c^{(\prime)}\bar{D}_s^{*}$&$\Xi_c^{*}\bar{D}_s$&$\Xi_c^{*}\bar{D}^{*}_s$\\\midrule[1.0pt]
${1}/{2}^{-}$&$|{}^2\mathbb{S}_{1/2}\rangle$&$|{}^2\mathbb{S}_{1/2}\rangle/|{}^4\mathbb{D}_{1/2}\rangle$&$...$&$|{}^2\mathbb{S}_{1/2}\rangle/|{}^{4,6}\mathbb{D}_{1/2}\rangle$\\
${3}/{2}^{-}$ &$...$&$|{}^4\mathbb{S}_{3/2}\rangle/|{}^{2,4}\mathbb{D}_{3/2}\rangle$&$|{}^4\mathbb{S}_{3/2}\rangle/|{}^{4}\mathbb{D}_{3/2}\rangle$&$|{}^4\mathbb{S}_{3/2}\rangle/|{}^{2,4,6}\mathbb{D}_{3/2}\rangle$\\
${5}/{2}^{-}$&$...$&$...$&$...$&$|{}^6\mathbb{S}_{5/2}\rangle/|{}^{2,4,6}\mathbb{D}_{5/2}\rangle$\\
\bottomrule[1pt]\bottomrule[1pt]
\end{tabular}
\end{table}

In addition, the normalization relations for the heavy hadrons $D_{s}$, $D_{s}^{*}$, $\mathcal{B}_{c}$, and $\mathcal{B}_{c}^{*}$ can be written as
\begin{eqnarray}
\langle 0|D_{s}|c\bar{s}\left(0^-\right)\rangle&=&\sqrt{M_{D_{s}}},\\
\langle 0|D_{s}^{*\mu}|c\bar{s}\left(1^-\right)\rangle&=&\sqrt{M_{D_{s}^*}}\epsilon^\mu,\\
\langle 0|\mathcal{B}_{c}|cqq\left({1}/{2}^+\right)\rangle &=& \sqrt{2M_{\mathcal{B}_{c}}}{\left(\chi_{\frac{1}{2}m},\frac{\bf{\sigma}\cdot\bf{p}}{2M_{\mathcal{B}_{c}}}\chi_{\frac{1}{2},m}\right)^T},\\
\langle 0|\mathcal{B}_{c}^{*\mu}|cqq\left({3}/{2}^+\right)\rangle &=&\sum_{m_1,m_2}C_{1/2,m_1;1,m_2}^{3/2,m_1+m_2}\sqrt{2M_{\mathcal{B}_{c}^*}}\nonumber\\
  &&\times\left(\chi_{\frac{1}{2},m_1},\frac{\bf{\sigma}\cdot\bf{p}}{2M_{\mathcal{B}_c^*}}\chi_{\frac{1}{2}m_1}\right)^T\epsilon^{\mu}_{m_2},
\end{eqnarray}
respectively. Here, $M_{a}$ ($a=D_{s},~D_{s}^{*},~\mathcal{B}_{c},~\mathcal{B}_{c}^*$) denotes the corresponding mass of the heavy hadrons, and $\bm{p}$ is the momentum of the corresponding heavy hadrons.

Through the above three typical steps, we can derive the effective potentials in the coordinate space for all of the investigated systems, where these obtained effective potentials are a little more complicated, which are collected in Appendix~\ref{app01}.

\section{Finding bound state solution of the investigated systems}\label{sec3}
In our calculation, we need a series of input parameters of the coupling constants. These coupling constants can be extracted from the experimental data or by the theoretical model, and the corresponding phase factors between these coupling constants can be fixed by the quark model \cite{Riska:2000gd}. The values of these involved parameters include $g_S=0.76$, $l_B=-3.65$, $l_S=6.20$, $g=0.59$, $g_4=1.06$, $g_1=0.94$, $f_\pi=132~\rm{MeV}$, $\beta g_V=-5.25$, $\beta_B g_V=-6.00$, $\beta_S g_V=10.14$, $\lambda g_V =-3.27~\rm{GeV}^{-1}$, $\lambda_I g_V =-6.80~\rm{GeV}^{-1}$, and $\lambda_S g_V=19.2~\rm{GeV}^{-1}$ \cite{Chen:2017xat,Chen:2019asm}.  In addition, the adopted hadron masses are $m_\sigma=600.00$ MeV, $m_\eta=547.85$ MeV, $m_\phi=1019.46$ MeV, $m_{D_s}=1968.34$ MeV, $m_{D_s^*}=2112.20$ MeV, $m_{\Xi_{c}}=2469.42$ MeV, $m_{\Xi_{c}^{\prime}}=2578.80$ MeV, and $m_{\Xi_{c}^*}=2645.97$ MeV \cite{Zyla:2020zbs}.

After getting the general expressions of the effective potentials for these discussed systems, we solve the coupled channel Schr$\rm{\ddot{o}}$dinger equation and attempt to find the bound state solution by varying the cutoff parameter $\Lambda$, i.e.,
\begin{eqnarray}\label{SE}
-\frac{1}{2\mu}\left(\nabla^2-\frac{\ell(\ell+1)}{r^2}\right)\psi(r)+V(r)\psi(r)=E\psi(r)
\end{eqnarray}
with $\nabla^2=\frac{1}{r^2}\frac{\partial}{\partial r}r^2\frac{\partial}{\partial r}$, where $\mu=\frac{m_1m_2}{m_1+m_2}$ is the reduced mass for the investigated system. According to Eq. (\ref{SE}), we find that there exists the repulsive centrifugal potential $\ell(\ell+1)/2\mu r^2$ for the higher partial wave states. Thus, we are mainly interested in the $S$-wave $\Xi_c^{(\prime,*)}\bar{D}_s^{(*)}$ systems in this work.

Before producing the numerical calculation, we firstly analyze the properties of the OBE effective potentials for the $\Xi_c^{(\prime,*)}\bar{D}_s^{(*)}$ systems. For the $S$-wave $\Xi_c\bar D_s$, $\Xi_c^{\prime}\bar D_s$, $\Xi_c\bar D_s^{*}$, and $\Xi_c^{*}\bar D_s$ systems, the $\sigma$ and $\phi$ exchanges contribute to the total effective potentials. And, there exists the $\sigma$, $\eta$, and $\phi$ exchange interactions for the $S$-wave $\Xi_c^{\prime}\bar D_s^*$ and $\Xi_c^{*}\bar D_s^*$ systems.

In our numerical analysis, we will discuss the bound state properties of these investigated systems by performing single channel analysis and coupled channel analysis, and the detailed information of how to consider the $S$-$D$ wave mixing effect and the coupled channel effect can be referred to Ref. \cite{Wang:2019nwt}. 

\subsection{The $\Xi_{c}^{\prime}\bar D_s$ system}

For the $S$-wave $\Xi_{c}^{\prime}\bar D_s$ state with $J^P={1}/{2}^{-}$, we cannot find the bound state solution corresponding to the cutoff $\Lambda$ range from 1.00 GeV to 5.00 GeV when taking the single channel analysis.
If further considering the coupled channel effect from the $\Xi_{c}^{\prime}\bar D_s|{}^2\mathbb{S}_{1/2}\rangle$, $\Xi_{c}\bar D_s^*|{}^2\mathbb{S}_{1/2}\rangle$, $\Xi_{c}^{\prime}\bar D_s^*|{}^2\mathbb{S}_{1/2}\rangle$, and $\Xi_{c}^{*}\bar D_s^*|{}^2\mathbb{S}_{1/2}\rangle$ channels, the numerical result is shown in Table~\ref{r1}, where the obtained bound state solution for the $S$-wave $\Xi_{c}^{\prime}\bar D_s$ state with $J^P={1}/{2}^{-}$ is given.
\renewcommand\tabcolsep{0.37cm}
\renewcommand{\arraystretch}{1.50}
\begin{table}[!htbp]
\caption{Bound state properties for the $S$-wave $\Xi_{c}^{\prime}\bar D_s$ state with $J^P={1}/{2}^{-}$ by performing the coupled channel analysis. Here, the cutoff $\Lambda$, binding energy $E$, and root-mean-square radius $r_{RMS}$ are in units of $ \rm{GeV}$, $\rm {MeV}$, and $\rm {fm}$, respectively.}\label{r1}
\begin{tabular}{cccc}\toprule[1pt]\toprule[1pt]
$\Lambda$ &$E$  &$r_{\rm RMS}$ &P($\Xi_{c}^{\prime}\bar D_s/\Xi_{c}\bar D_s^*/\Xi_{c}^{\prime}\bar D_s^*/\Xi_{c}^{*}\bar D_s^*$)\\
\cline{1-4}
2.44&$-1.05$ &2.81&\textbf{92.39}/4.67/2.04/0.91\\
2.46&$-7.80$ &0.99&\textbf{80.70}/11.86/5.15/2.29\\
2.47&$-13.19$ &0.75&\textbf{76.38}/14.44/6.35/2.83\\
\bottomrule[1pt]\bottomrule[1pt]
\end{tabular}
\end{table}

For the $S$-wave $\Xi_{c}^{\prime}\bar D_s$ state with $J^P={1}/{2}^{-}$, we find that the binding energy can reach up to several MeV when taking the cutoff parameter $\Lambda$ to be around 2.45 GeV, where the $\Xi_{c}^{\prime}\bar D_s$ component is dominant. Thus, our study indicates that the contribution from the coupled channel effect may play an important role for forming the $S$-wave $\Xi_{c}^{\prime}\bar D_s$ bound state with $J^P={1}/{2}^{-}$ \cite{Chen:2019asm}.

\subsection{The $\Xi_{c}\bar D_s^*$ system}
For the $S$-wave $\Xi_{c}\bar D_s^*$ state with $J^P={1}/{2}^{-}$, we cannot find the bound state solution with the single channel analysis when scanning the cutoff parameter range $\Lambda=1.00$-$5.00~{\rm GeV}$. As shown in Table~\ref{r4}, if we consider the coupled channel effect and take the cutoff parameter $\Lambda$ around 4.31 GeV, we can obtain bound state solution for the $S$-wave $\Xi_{c}\bar D_s^*$ state with $J^P={1}/{2}^{-}$, and the $\Xi_{c}\bar D_s^*$ system is the dominant channel with probabilities over 90\%. However, the corresponding cutoff is far from the usual value around 1.00 GeV \cite{Tornqvist:1993ng,Tornqvist:1993vu}. Thus, we cannot recommend the $S$-wave $\Xi_{c}\bar D_s^*$ state with $J^P={1}/{2}^{-}$ as prior candidate of molecular state.
\renewcommand\tabcolsep{0.47cm}
\renewcommand{\arraystretch}{1.50}
\begin{table}[!htbp]
\caption{Bound state properties for the $S$-wave $\Xi_{c}\bar D_s^*$ state with $J^P={1}/{2}^{-}$ when the coupled channel effect is introduced.}\label{r4}
\begin{tabular}{cccc}\toprule[1pt]\toprule[1pt]
$\Lambda$ &$E$  &$r_{\rm RMS}$ &P($\Xi_{c}\bar D_s^*/\Xi_{c}^{\prime}\bar D_s^*/\Xi_{c}^{*}\bar D_s^*$)\\
\cline{1-4}
4.31&$-0.24$ &4.84&\textbf{99.25}/0.50/0.25\\
4.34&$-4.02$ &1.46&\textbf{97.03}/2.00/0.97\\
4.37&$-12.54$ &0.80&\textbf{94.78}/3.52/1.70\\
\bottomrule[1pt]\bottomrule[1pt]
\end{tabular}
\end{table}

For the $S$-wave $\Xi_{c}\bar D_s^*$ state with $J^P={3}/{2}^{-}$, we cannot find the {reasonable} bound state solution corresponding to $\Lambda=1.00\sim 5.00$ GeV, even if we consider the contribution of the coupled channel effect. {{Here, we need to point out that we can find bound state solutions for the coupled $\Xi_{c}\bar D_s^*/\Xi_{c}^{\prime}\bar D_s^*/\Xi_{c}^{*}\bar D_s^*$ system with $J^P={3}/{2}^{-}$ when considering the coupled channel effect. However, the dominant channel is not the system with lowest mass threshold, the corresponding RMS radius is often smaller than 0.5 fm \cite{Chen:2017xat}. Obviously, it cannot be a loosely bound hadronic molecular candidate. We also perform the same analysis in the following discussions.}} Thus, our result disfavors the existence of  the $S$-wave $\Xi_{c}\bar D_s^*$ bound state with $J^P={3}/{2}^{-}$.

\subsection{The $\Xi_{c}^{\prime}\bar D_s^*$ system}

With single channel analysis and coupled channel analysis, there does not exist the reasonable bound state solution for the $S$-wave $\Xi_{c}^{\prime}\bar D_s^*$ state with $J^P={1}/{2}^{-}$.
\renewcommand\tabcolsep{0.32cm}
\renewcommand{\arraystretch}{1.50}
\begin{table}[!htbp]
\caption{Bound state properties for the $S$-wave $\Xi_{c}^{\prime}\bar D_s^*$ state with $J^P={3}/{2}^{-}$.  Here, we listed the numerical result with the single channel analysis and including the $S$-$D$ mixing effect, and the coupled channel effect which correspond to Case-I, Case-II, and Case-III, respectively.}\label{r2}
\begin{tabular}{ccccc}\toprule[1pt]\toprule[1pt]
&$\Lambda$ &$E$  &$r_{\rm RMS}$ &\\
\hline
&2.52&$-0.52$ &3.81&  \\
Case-I&2.57&$-5.17$ &1.28&      \\
&2.61&$-12.38$ &0.83&      \\
\hline
&$\Lambda$ &$E$  &$r_{\rm RMS}$ &P(${}^4\mathbb{S}_{\frac{3}{2}}/{}^2\mathbb{D}_{\frac{3}{2}}/{}^4\mathbb{D}_{\frac{3}{2}})$\\\hline
&2.45&$-0.43$ &4.09&\textbf{99.83}/0.02/0.14\\
Case-II&2.50&$-4.73$ &1.35&\textbf{99.50}/0.08/0.42\\
&2.54&$-11.55$ &0.87&\textbf{99.28}/0.11/0.61\\
\hline
&$\Lambda$ &$E$  &$r_{\rm RMS}$ &P($\Xi_{c}^{\prime}\bar D_s^*/\Xi_{c}^{*}\bar D_s^*$)\\
\hline
&2.09&$-0.22$ &4.91&\textbf{98.93}/1.07\\
Case-III&2.13&$-3.93$ &1.47&\textbf{95.81}/4.19\\
&2.17&$-12.23$ &0.83&\textbf{93.00}/7.00\\
\bottomrule[1pt]\bottomrule[1pt]
\end{tabular}
\end{table}

For the $S$-wave $\Xi_{c}^{\prime}\bar D_s^*$ state with $J^P={3}/{2}^{-}$, we list the numerical results of finding bound state solution in Table~\ref{r2}. The loosely binding energies can be obtained under three cases. With including the $S$-$D$ mixing effect and the coupled channel effect step by step, the cutoff value becomes smaller if getting the same binding energy. Although the coupled channel effect is included in our study, we find that the $\Xi_{c}^{\prime}\bar D_s^*$ channel contribution is dominant. Generally speaking, the coupled channel effect is propitious to form bound state for the $\Xi_{c}^{\prime}\bar D_s^*$ state with $J^P={3}/{2}^{-}$.

\subsection{The $\Xi_{c}^{*}\bar D_s^*$ system}
\renewcommand\tabcolsep{0.05cm}
\renewcommand{\arraystretch}{1.50}
\begin{table}[!htbp]
\caption{Bound state properties for the $S$-wave $\Xi_{c}^{*}\bar D_s^*$ system. The second column shows the obtained bound state solutions without taking into account the $S$-$D$ wave mixing effect, and the last column is the relevant result by performing the $S$-$D$ wave mixing effect analysis. Here, ``$\times$" indicates no binding energy when scanning the cutoff parameter $\Lambda$ range from 1.00 GeV to 5.00 GeV.}\label{r3}
\begin{tabular}{c|ccc|cccc}\toprule[1pt]\toprule[1pt]
$J^P$&$\Lambda$ &$E$  &$r_{\rm RMS}$ &$\Lambda$ &$E$  &$r_{\rm RMS}$ &P(${}^2\mathbb{S}_{\frac{1}{2}}/{}^4\mathbb{D}_{\frac{1}{2}}/{}^6\mathbb{D}_{\frac{1}{2}})$\\
\cline{1-8}
\multirow{3}{*}{${1}/{2}^{-}$}&$\times$&$\times$&$\times$&      4.31&$-0.27$ &4.93&\textbf{99.89}/0.07/0.04\\
                                  &$\times$&$\times$&$\times$&      4.65&$-1.71$ &2.50&\textbf{99.69}/0.21/0.10\\
                                  &$\times$&$\times$&$\times$&      5.00&$-4.60$ &1.63&\textbf{99.41}/0.40/0.19\\
\cline{1-8}
$J^P$&$\Lambda$ &$E$  &$r_{\rm RMS}$ &$\Lambda$ &$E$  &$r_{\rm RMS}$ &P(${}^6\mathbb{S}_{\frac{5}{2}}/{}^2\mathbb{D}_{\frac{5}{2}}/{}^4\mathbb{D}_{\frac{5}{2}}/{}^6\mathbb{D}_{\frac{5}{2}})$\\
\cline{1-8}
\multirow{3}{*}{${5}/{2}^{-}$}&2.05&$-0.36$ &4.31&      2.03&$-0.28$ &4.63&\textbf{99.95}/$o(0)$/$o(0)$/0.05\\
                                  &2.10&$-5.09$ &1.30&      2.08&$-4.86$ &1.34&\textbf{99.79}/0.02/0.01/0.18\\
                                  &2.14&$-13.03$ &0.82&     2.12&$-12.74$ &0.84&\textbf{99.68}/0.02/0.01/0.29\\
\bottomrule[1pt]\bottomrule[1pt]
\end{tabular}
\end{table}

For the $S$-wave $\Xi_{c}^{*}\bar D_s^*$ system, the relevant bound state properties are collected in Table~\ref{r3}. In the following, we summarize several main points:
\begin{itemize}
  \item For the $J^P=1/2^{-}$ system, we cannot find bound state solution with single channel analysis. After considering the $S$-$D$ wave mixing effect, we may find shallow binding energy when taking the cutoff parameter $\Lambda$ around 4.31 GeV, which in fact is deviate from 1.00 GeV \cite{Tornqvist:1993ng,Tornqvist:1993vu}. Thus, our result disfavors the existence of the $S$-wave $\Xi_{c}^{*}\bar D_s^*$ molecular state with $J^P={1}/{2}^{-}$.
  \item For the $J^P=3/2^{-}$ system, the bound state solution cannot be found even if the contribution of the $S$-$D$ wave mixing effect is introduced. Therefore, we may conclude that the $\Xi_{c}^{*}\bar D_s^*$ state with $J^P={3}/{2}^{-}$ cannot be bound together as a molecular state candidate.
  \item For the $J^P=5/2^{-}$ system, when the cutoff parameter $\Lambda$ is slightly bigger than 2.00 GeV, this system exists the bound state solution with small binding energies and suitable RMS radius by performing single channel analysis. By considering the $S$-$D$ wave mixing effect in our calculation, we find that the $\Xi_{c}^{*}\bar D_s^*|{}^6\mathbb{S}_{5/2}\rangle$ channel has the dominant contribution with the probability over 99 percent and plays a major role for forming the $S$-wave $\Xi_{c}^{*}\bar D_s^*$ bound state with $J^P={5}/{2}^{-}$ \cite{Wang:2019nwt}.
  \end{itemize}

In this work, we also study the $S$-wave $\Xi_{c}\bar D_s$ state with $J^P={1}/{2}^{-}$ and the $S$-wave $\Xi_{c}^*\bar D_s$ state with $J^P={3}/{2}^{-}$. When all effects presented in this work, we cannot find the corresponding reasonable bound state solution when tuning cutoff parameter $\Lambda$  from 1.00 GeV to 5.00 GeV. Thus, the $S$-wave $\Xi_{c}\bar D_s$ state with $J^P={1}/{2}^{-}$ and the $S$-wave $\Xi_{c}^*\bar D_s$ state with $J^P={3}/{2}^{-}$ cannot form hadronic molecular states.

{Compared to the $P_c$ and $P_{cs}$ states assigned to meson-baryon molecules, the long rang interaction from the one $\pi$ exchange is absent for the $\Xi_c^{(',*)}\bar{D}_s^{(*)}$ systems. This explains that a larger cutoff value is needed to produce a bound solution. In general, a bound state with a smaller cutoff input is more likely to recommend as a molecular candidate. In this evaluation, we finally conclude the $S$-wave $\Xi_{c}^{*}\bar D_s^*$ molecular state with $J^P={5}/{2}^{-}$ and the $\Xi_{c}^{\prime}\bar D_s^*$ molecular state with $J^P={3}/{2}^{-}$ may be assigned as hidden-charm molecular pentaquarks.}

\section{Summary}\label{sec4}

Very recently, the LHCb Collaboration reported the evidence of a new $P_{cs}(4459)$ structure existing in the $J/\psi \Lambda$ invariant mass spectrum when analyzing the $\Xi_b^-\to J/\psi K^- \Lambda$ process. As the candidate of the $\Xi_c \bar{D}^*$ hadronic molecular state \cite{Chen:2020uif,Peng:2020hql,1830432,1830426,1830449}, $P_{cs}(4459)$ has strangeness $|S|=1$, which makes hidden-charm pentaquark family can be extended by adding strangeness.
Along this line, it is natural to focus on hidden-charm molecular pentaquarks with double strangeness, which is the main task of the present work.

In this work, we study the bound state properties of the systems composed of an $S$-wave charmed baryon $\Xi_c^{(\prime,*)}$ and an $S$-wave anti-charmed-strange meson $\bar{D}_s^{(*)}$, where the OBE model is applied in our concrete calculation. By carrying out a quantitative calculation, we suggest that there may exist the candidates of hidden-charm pentaquark with double strangeness, which include the $S$-wave $\Xi_{c}^{*}\bar D_s^*$ molecular state with $J^P={5}/{2}^{-}$ and the $\Xi_{c}^{\prime}\bar D_s^*$ molecular state with $J^P={3}/{2}^{-}$. Experimental searches for them will be a challenge and an opportunity, especially for LHCb.

\begin{figure}[!htbp]
\includegraphics[width=5.50cm,keepaspectratio]{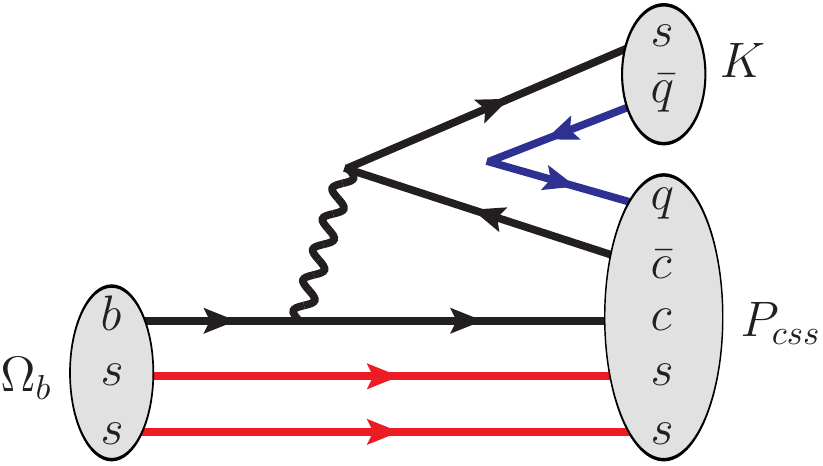}
\caption{The production process involved in the predicted hidden-charm pentaquarks with double strangeness.}
\label{fig2}
\end{figure}

We may further discuss the production mechanism for them in the following. As shown in Fig. \ref{fig2}, it is highly probable that these predicted hidden-charm pentaquarks with double strangeness can be detected through the $\Omega_b$ baryon weak decays\footnote{The hidden-charm pentaquark with double strangeness is marked as the $P_{css}$ state in Fig. \ref{fig2}.} in the future, since the fact that three $P_c$ states and the $P_{cs}(4459)$ structure are from the $\Lambda_b$ baryon weak decay \cite{Aaij:2019vzc} and the $\Xi_b$ baryon weak decay \cite{lhcb}, respectively. Therefore, we hope that the future experiment like LHCb can bring us more surprises when analyzing the $J/\psi \Xi$ invariant mass spectrum of the $\Omega_b \to J/\psi \Xi K$ process. Obviously, it will crucial step when constructing hidden-charm pentaquark zoo.

\section*{ACKNOWLEDGMENTS}
This work is supported by the China National Funds for Distinguished Young Scientists under Grant No. 11825503, the National Program for Support of Top-notch Young Professionals and the 111 Project under Grant No. B20063. R. C. is supported by the National Postdoctoral Program for Innovative Talent.

\appendix
\section{Relevant subpotentials}\label{app01}
The effective potentials for these discussed processes are expressed as
\begin{eqnarray}
\mathcal{V}^{\Xi_{c}\bar D_s\rightarrow\Xi_{c}\bar D_s}&=&2AY_\sigma+\frac{E}{2}Y_{\phi},\\
\mathcal{V}^{\Xi_{c}\bar D_s\rightarrow\Xi_{c}^{\prime}\bar D_s^*}&=&-\frac{C}{3\sqrt{6}}\left[\mathcal{A}_1\mathcal{O}_r+\mathcal{A}_2\mathcal{P}_r\right]Y_{\eta1}\nonumber\\
&&+\frac{2F}{3\sqrt{6}}\left[2\mathcal{A}_1\mathcal{O}_r-\mathcal{A}_2\mathcal{P}_r\right]Y_{\phi1},\\
\mathcal{V}^{\Xi_{c}\bar D_s\rightarrow\Xi_{c}^*\bar D_s^*}&=&\frac{C}{3\sqrt{2}}\left[\mathcal{A}_3\mathcal{O}_r+\mathcal{A}_4\mathcal{P}_r\right]Y_{\eta2}\nonumber\\
&&-\frac{\sqrt{2}F}{3}\left[2\mathcal{A}_3\mathcal{O}_r-\mathcal{A}_4\mathcal{P}_r\right]Y_{\phi2},\\
\mathcal{V}^{\Xi_{c}^{\prime}\bar D_s\rightarrow\Xi_{c}^{\prime}\bar D_s}&=&-BY_\sigma-\frac{G}{4}Y_{\phi},\\
\mathcal{V}^{\Xi_{c}^{\prime}\bar D_s\rightarrow\Xi_{c}\bar D_s^*}&=&-\frac{C}{3\sqrt{6}}\left[\mathcal{A}_1\mathcal{O}_r+\mathcal{A}_2\mathcal{P}_r\right]Y_{\eta3}\nonumber\\
&&+\frac{2F}{3\sqrt{6}}\left[2\mathcal{A}_1\mathcal{O}_r-\mathcal{A}_2\mathcal{P}_r\right]Y_{\phi3},\\
\mathcal{V}^{\Xi_{c}^{\prime}\bar D_s\rightarrow\Xi_{c}^*\bar D_s}&=&\frac{B}{\sqrt{3}}\mathcal{A}_5Y_{\sigma4}+\frac{G}{4\sqrt{3}}\mathcal{A}_5Y_{\phi4},\\
\mathcal{V}^{\Xi_{c}^{\prime}\bar D_s\rightarrow\Xi_{c}^{\prime}\bar D_s^*}&=&\frac{D}{18}\left[\mathcal{A}_1\mathcal{O}_r+\mathcal{A}_2\mathcal{P}_r\right]Y_{\eta5}\nonumber\\
&&+\frac{H}{9}\left[2\mathcal{A}_1\mathcal{O}_r-\mathcal{A}_2\mathcal{P}_r\right]Y_{\phi5},\\
\mathcal{V}^{\Xi_{c}^{\prime}\bar D_s\rightarrow\Xi_{c}^*\bar D_s^*}&=&-\frac{D}{12\sqrt{3}}\left[\mathcal{A}_6\mathcal{O}_r+\mathcal{A}_7\mathcal{P}_r\right]Y_{\eta6}\nonumber\\
&&-\frac{H}{6\sqrt{3}}\left[2\mathcal{A}_6\mathcal{O}_r-\mathcal{A}_7\mathcal{P}_r\right]Y_{\phi6},\\
\mathcal{V}^{\Xi_{c}\bar D_s^*\rightarrow\Xi_{c}\bar D_s^*}&=&2A\mathcal{A}_{8}Y_\sigma+\frac{E}{2}\mathcal{A}_{8}Y_{\phi},\\
\mathcal{V}^{\Xi_{c}\bar D_s^*\rightarrow\Xi_{c}^*\bar D_s}&=&\frac{C}{3\sqrt{2}}\left[\mathcal{A}_{9}\mathcal{O}_r+\mathcal{A}_{10}\mathcal{P}_r\right]Y_{\eta7}\nonumber\\
&&-\frac{\sqrt{2}F}{3}\left[2\mathcal{A}_{9}\mathcal{O}_r-\mathcal{A}_{10}\mathcal{P}_r\right]Y_{\phi7},\\
\mathcal{V}^{\Xi_{c}\bar D_s^*\rightarrow\Xi_{c}^{\prime}\bar D_s^*}&=&-\frac{C}{3\sqrt{6}}\left[\mathcal{A}_{11}\mathcal{O}_r+\mathcal{A}_{12}\mathcal{P}_r\right]Y_{\eta8}\nonumber\\
&&+\frac{2F}{3\sqrt{6}}\left[2\mathcal{A}_{11}\mathcal{O}_r-\mathcal{A}_{12}\mathcal{P}_r\right]Y_{\phi8},\\
\mathcal{V}^{\Xi_{c}\bar D_s^*\rightarrow\Xi_{c}^*\bar D_s^*}&=&\frac{C}{3\sqrt{2}}\left[\mathcal{A}_{13}\mathcal{O}_r+\mathcal{A}_{14}\mathcal{P}_r\right]Y_{\eta9}\nonumber\\
&&-\frac{\sqrt{2}F}{3}\left[2\mathcal{A}_{13}\mathcal{O}_r-\mathcal{A}_{14}\mathcal{P}_r\right]Y_{\phi9},\\
\mathcal{V}^{\Xi_{c}^*\bar D_s\rightarrow\Xi_{c}^*\bar D_s}&=&-B\mathcal{A}_{15}Y_\sigma-\frac{G}{4}\mathcal{A}_{15}Y_{\phi},\\
\mathcal{V}^{\Xi_{c}^*\bar D_s\rightarrow\Xi_{c}^{\prime}\bar D_s^*}&=&\frac{D}{12\sqrt{3}}\left[\mathcal{A}_{16}\mathcal{O}_r+\mathcal{A}_{17}\mathcal{P}_r\right]Y_{\eta10}\nonumber\\
&&+\frac{H}{6\sqrt{3}}\left[2\mathcal{A}_{16}\mathcal{O}_r-\mathcal{A}_{17}\mathcal{P}_r\right]Y_{\phi10},\\
\mathcal{V}^{\Xi_{c}^*\bar D_s\rightarrow\Xi_{c}^*\bar D_s^*}&=&\frac{D}{12}\left[\mathcal{A}_{18}\mathcal{O}_r+\mathcal{A}_{19}\mathcal{P}_r\right]Y_{\eta11}\nonumber\\
&&+\frac{H}{6}\left[2\mathcal{A}_{18}\mathcal{O}_r-\mathcal{A}_{19}\mathcal{P}_r\right]Y_{\phi11},\\
\mathcal{V}^{\Xi_{c}^{\prime}\bar D_s^*\rightarrow\Xi_{c}^{\prime}\bar D_s^*}&=&-B\mathcal{A}_{8}Y_\sigma+\frac{D}{18}\left[\mathcal{A}_{11}\mathcal{O}_r+\mathcal{A}_{12}\mathcal{P}_r\right]Y_{\eta}\nonumber\\
&&-\frac{G}{4}\mathcal{A}_{8}Y_{\phi}-\frac{H}{9}\left[2\mathcal{A}_{11}\mathcal{O}_r-\mathcal{A}_{12}\mathcal{P}_r\right]Y_{\phi},\\
\mathcal{V}^{\Xi_{c}^{\prime}\bar D_s^*\rightarrow\Xi_{c}^*\bar D_s^*}&=&\frac{B}{\sqrt{3}}\mathcal{A}_{20}Y_{\sigma12}+\frac{D}{12\sqrt{3}}\left[\mathcal{A}_{21}\mathcal{O}_r+\mathcal{A}_{22}\mathcal{P}_r\right]Y_{\eta12}\nonumber\\
&&+\frac{G\mathcal{A}_{20}}{4\sqrt{3}}Y_{\phi12}-\frac{H}{6\sqrt{3}}\left[2\mathcal{A}_{21}\mathcal{O}_r-\mathcal{A}_{22}\mathcal{P}_r\right]Y_{\phi12},\nonumber\\\\
\mathcal{V}^{\Xi_{c}^*\bar D_s^*\rightarrow\Xi_{c}^{*}\bar D_s^*}&=&-B\mathcal{A}_{23}Y_\sigma-\frac{D}{12}\left[\mathcal{A}_{24}\mathcal{O}_r+\mathcal{A}_{25}\mathcal{P}_r\right]Y_{\eta}\nonumber\\
&&-\frac{G}{4}\mathcal{A}_{23}Y_{\phi}+\frac{H}{6}\left[2\mathcal{A}_{24}\mathcal{O}_r-\mathcal{A}_{25}\mathcal{P}_r\right]Y_{\phi}.
\end{eqnarray}
Here, $\mathcal{O}_r = \frac{1}{r^2}\frac{\partial}{\partial r}r^2\frac{\partial}{\partial r}$ and $\mathcal{P}_r = r\frac{\partial}{\partial r}\frac{1}{r}\frac{\partial}{\partial r}$. Additionally, we also define several variables, which include $A=l_{B}g_S$, $B=l_Sg_S$, $C=g_4 g/f_\pi^2$, $D=g_1 g/f_\pi^2$, $E=\beta \beta_B g_{V}^2$, $F=\lambda \lambda_Ig_V^2$, $G=\beta \beta_S g_{V}^2$, and $H=\lambda \lambda_S g_V^2$. The function $Y_i$ is defined as
\begin{eqnarray}
Y_i\equiv \dfrac{e^{-m_ir}-e^{-\Lambda_ir}}{4\pi r}-\dfrac{\Lambda_i^2-m_i^2}{8\pi\Lambda_i}e^{-\Lambda_ir}.
\end{eqnarray}
Here, $m_i=\sqrt{m^2-q_i^2}$ and $\Lambda_i=\sqrt{\Lambda^2-q_i^2}$ with $q_i^2 =(m_A^2+m_D^2-m_B^2-m_C^2)^2/(2m_C+2m_D)^2$ for the processes $AB \to CD$.

In the above OBE effective potentials, we also introduce several operators $\mathcal{A}_k$, i.e.,
\begin{eqnarray*}
\mathcal{A}_{1}&=&\chi^{\dagger}_3\left({\bm\sigma}\cdot{\bm\epsilon^{\dagger}_{4}}\right)\chi_1,\nonumber\\
\mathcal{A}_{2}&=&\chi^{\dagger}_3T({\bm\sigma},{\bm\epsilon^{\dagger}_{4}})\chi_1,\nonumber\\
\mathcal{A}_{3}&=&\sum_{a,b}C^{\frac{3}{2},a+b}_{\frac{1}{2}a,1b}\chi^{\dagger a}_{3}\left({\bm\epsilon^{\dagger b}_{3}}\cdot{\bm\epsilon^{\dagger}_{4}}\right)\chi_1,\nonumber\\
\mathcal{A}_{4}&=&\sum_{a,b}C^{\frac{3}{2},a+b}_{\frac{1}{2}a,1b}\chi^{\dagger a}_{3}T({\bm\epsilon^{\dagger b}_{3}},{\bm\epsilon^{\dagger}_{4}})\chi_1,\nonumber\\
\mathcal{A}_{5}&=&\sum_{a,b}C^{\frac{3}{2},a+b}_{\frac{1}{2}a,1b}\chi^{\dagger a}_{3}\left({\bm\epsilon^{\dagger b}_{3}}\cdot{\bm\sigma}\right)\chi_1,\nonumber\\
\mathcal{A}_{6}&=&\sum_{a,b}C^{\frac{3}{2},a+b}_{\frac{1}{2}a,1b}\chi^{\dagger a}_3\left[{\bm\epsilon^{\dagger}_{4}}\cdot\left(i{\bm\sigma}\times{\bm\epsilon^{\dagger b}_{3}}\right)\right]\chi_1,\nonumber\\
\mathcal{A}_{7}&=&\sum_{a,b}C^{\frac{3}{2},a+b}_{\frac{1}{2}a,1b}\chi^{\dagger a}_3T({\bm\epsilon^{\dagger}_{4}},i{\bm\sigma}\times{\bm\epsilon^{\dagger b}_{3}})\chi_1,\nonumber\\
\mathcal{A}_{8}&=&\chi^{\dagger}_3\left({\bm\epsilon^{\dagger}_{4}}\cdot{\bm\epsilon_{2}}\right)\chi_1,\nonumber\\
\mathcal{A}_{9}&=&\sum_{a,b}C^{\frac{3}{2},a+b}_{\frac{1}{2}a,1b}\chi^{\dagger a}_{3}\left({\bm\epsilon^{\dagger b}_{3}}\cdot{\bm\epsilon_{2}}\right)\chi_1,\nonumber\\
\mathcal{A}_{10}&=&\sum_{a,b}C^{\frac{3}{2},a+b}_{\frac{1}{2}a,1b}\chi^{\dagger a}_{3}T({\bm\epsilon^{\dagger b}_{3}},{\bm\epsilon_{2}})\chi_1,\nonumber\\
\mathcal{A}_{11}&=&\chi^{\dagger}_3\left[{\bm\sigma}\cdot\left(i{\bm\epsilon_{2}}\times{\bm\epsilon^{\dagger}_{4}}\right)\right]\chi_1,\nonumber\\
\mathcal{A}_{12}&=&\chi^{\dagger}_3T({\bm\sigma},i{\bm\epsilon_{2}}\times{\bm\epsilon^{\dagger}_{4}})\chi_1,\nonumber\\
\mathcal{A}_{13}&=&\sum_{a,b}C^{\frac{3}{2},a+b}_{\frac{1}{2}a,1b}\chi^{\dagger a}_3\left[{\bm\epsilon^{\dagger b}_{3}}\cdot\left(i{\bm\epsilon_{2}}\times{\bm\epsilon^{\dagger}_{4}}\right)\right]\chi_1,\nonumber\\
\mathcal{A}_{14}&=&\sum_{a,b}C^{\frac{3}{2},a+b}_{\frac{1}{2}a,1b}\chi^{\dagger a}_3T({\bm\epsilon^{\dagger b}_{3}},i{\bm\epsilon_{2}}\times{\bm\epsilon^{\dagger}_{4}})\chi_1,\nonumber\\
\mathcal{A}_{15}&=&\sum_{a,b,m,n}C^{\frac{3}{2},a+b}_{\frac{1}{2}a,1b}C^{\frac{3}{2},m+n}_{\frac{1}{2}m,1n}\chi^{\dagger a}_{3}\left({\bm\epsilon^{\dagger b}_{3}}\cdot{\bm\epsilon^{n}_{1}}\right)\chi^{m}_1,\nonumber\\
\mathcal{A}_{16}&=&\sum_{a,b}C^{\frac{3}{2},a+b}_{\frac{1}{2}a,1b}\chi^{\dagger}_3\left[{\bm\epsilon^{\dagger}_{4}}\cdot\left(i{\bm\sigma}\times{\bm\epsilon^{b}_{1}}\right)\right]\chi^a_1,\nonumber\\
\mathcal{A}_{17}&=&\sum_{a,b}C^{\frac{3}{2},a+b}_{\frac{1}{2}a,1b}\chi^{\dagger}_3T({\bm\epsilon^{\dagger}_{4}},i{\bm\sigma}\times{\bm\epsilon^{b}_{1}})\chi^a_1,\nonumber\\
\mathcal{A}_{18}&=&\sum_{a,b,m,n}C^{\frac{3}{2},a+b}_{\frac{1}{2}a,1b}C^{\frac{3}{2},m+n}_{\frac{1}{2}m,1n}\chi^{\dagger a}_3\left[{\bm\epsilon^{\dagger}_{4}}\cdot\left(i{\bm\epsilon^n_{1}}\times{\bm\epsilon^{\dagger b}_{3}}\right)\right]\chi^m_1,\nonumber\\
\mathcal{A}_{19}&=&\sum_{a,b,m,n}C^{\frac{3}{2},a+b}_{\frac{1}{2}a,1b}C^{\frac{3}{2},m+n}_{\frac{1}{2}m,1n}\chi^{\dagger a}_3T({\bm\epsilon^{\dagger}_{4}},i{\bm\epsilon^n_{1}}\times{\bm\epsilon^{\dagger b}_{3}})\chi^m_1,\nonumber\\
\mathcal{A}_{20}&=&\sum_{a,b}C^{\frac{3}{2},a+b}_{\frac{1}{2}a,1b}\chi^{\dagger a}_3\left({\bm\sigma}\cdot{\bm\epsilon^{\dagger b}_{3}}\right)\left({\bm\epsilon_{2}}\cdot{\bm\epsilon^{\dagger}_{4}}\right)\chi_1,\nonumber\\
\mathcal{A}_{21}&=&\sum_{a,b}C^{\frac{3}{2},a+b}_{\frac{1}{2}a,1b}\chi^{\dagger a}_3\left({\bm\sigma}\times{\bm\epsilon^{\dagger b}_{3}}\right)\cdot\left({\bm\epsilon_{2}}\times{\bm\epsilon^{\dagger}_{4}}\right)\chi_1,\nonumber\\
\mathcal{A}_{22}&=&\sum_{a,b}C^{\frac{3}{2},a+b}_{\frac{1}{2}a,1b}\chi^{\dagger a}_3T({\bm\sigma}\times{\bm\epsilon^{\dagger b}_{3}},{\bm\epsilon_{2}}\times{\bm\epsilon^{\dagger}_{4}})\chi_1,\nonumber\\
\mathcal{A}_{23}&=&\sum_{a,b,m,n}C^{\frac{3}{2},a+b}_{\frac{1}{2}a,1b}C^{\frac{3}{2},m+n}_{\frac{1}{2}m,1n}\chi^{\dagger a}_3\left({\bm\epsilon^{n}_{1}}\cdot{\bm\epsilon^{\dagger b}_{3}}\right)\left({\bm\epsilon_{2}}\cdot{\bm\epsilon^{\dagger}_{4}}\right)\chi^m_1,\nonumber\\
\mathcal{A}_{24}&=&\sum_{a,b,m,n}C^{\frac{3}{2},a+b}_{\frac{1}{2}a,1b}C^{\frac{3}{2},m+n}_{\frac{1}{2}m,1n}\chi^{\dagger a}_3\left({\bm\epsilon^{n}_{1}}\times{\bm\epsilon^{\dagger b}_{3}}\right)\cdot\left({\bm\epsilon_{2}}\times{\bm\epsilon^{\dagger}_{4}}\right)\chi^m_1,\nonumber\\
\mathcal{A}_{25}&=&\sum_{a,b,m,n}C^{\frac{3}{2},a+b}_{\frac{1}{2}a,1b}C^{\frac{3}{2},m+n}_{\frac{1}{2}m,1n}\chi^{\dagger a}_3T({\bm\epsilon^{n}_{1}}\times{\bm\epsilon^{\dagger b}_{3}},{\bm\epsilon_{2}}\times{\bm\epsilon^{\dagger}_{4}})\chi^m_1.\nonumber\\
\end{eqnarray*}
Here, the tensor force operator $T({\bm x},{\bm y})$ is expressed as $T({\bm x},{\bm y})= 3\left(\hat{\bm r} \cdot {\bm x}\right)\left(\hat{\bm r} \cdot {\bm y}\right)-{\bm x} \cdot {\bm y}$ with $\hat{\bm r}={\bm r}/|{\bm r}|$. In Table~\ref{matrix}, we present the obtained relevant matrix elements $\langle f|\mathcal{A}_k|i\rangle$, which will be used in our calculation.
\renewcommand\tabcolsep{0.22cm}
\renewcommand{\arraystretch}{1.70}
\begin{table*}[htbp]
  \caption{Matrix elements $\langle f|\mathcal{A}_k|i\rangle$ for the operators $\mathcal{A}_k$, which are obtained by sandwiching these operators between the relevant spin-orbit wave functions.}\label{matrix}
  \begin{tabular}{cccc|cc}\toprule[1pt]\toprule[1pt]
\multicolumn{4}{c|}{$S$-$D$ wave mixing effect}&\multicolumn{2}{c}{Coupled channel effect}\\ \midrule[1pt]
   {{{Spin}}} & $J=1/2$     & $J=3/2$     & $J=5/2$                     & $J=1/2$     & $J=3/2$\\\midrule[1pt]
 $\langle\mathcal{A}_{8}\rangle$
            &diag(1,1) &diag(1,1,1) &$/$        &$\mathcal{A}_{1}=\sqrt{3}$ &$\mathcal{A}_{9}=1$\\

 $\langle\mathcal{A}_{11}\rangle$
            &diag($-2$,$1$) &diag($1$,$-2$,$1$) &$/$          &$\mathcal{A}_{3}=-\sqrt{2}$ &$\mathcal{A}_{13}=-\sqrt{\frac{5}{3}}$\\

  $\langle\mathcal{A}_{12}\rangle$
           &$\left(\begin{array}{cc} 0 & -\sqrt{2} \\ -\sqrt{2} & -2\end{array}\right)$&$\left(\begin{array}{ccc} 0 & 1& 2 \\ 1 & 0& -1 \\ 2 & -1& 0 \end{array}\right)$&$/$           &$\begin{array}{l} \mathcal{A}_{6}=\sqrt{2}\\ \mathcal{A}_{13}=-\sqrt{\frac{2}{3}}\\ \mathcal{A}_{21}=-\sqrt{\frac{2}{3}}\end{array}$ &$\begin{array}{l} \mathcal{A}_{16}=1 \\ \mathcal{A}_{18}=\sqrt{\frac{5}{3}}\\ \mathcal{A}_{21}=-\sqrt{\frac{5}{3}} \end{array}$\\

 $\langle\mathcal{A}_{15}\rangle$
            &$/$ &diag(1,1) &$/$  & & \\

 $\langle\mathcal{A}_{23}\rangle$
            &diag(1,1,1) &diag(1,1,1,1) &diag(1,1,1,1)   &\\

 $\langle\mathcal{A}_{24}\rangle$
            &diag($\frac{5}{3}$,$\frac{2}{3}$,$-1$) &diag($\frac{2}{3}$,$\frac{5}{3}$,$\frac{2}{3}$,$-1$) &diag($-1$,$\frac{5}{3}$,$\frac{2}{3}$,$-1$)  \\

 $\langle\mathcal{A}_{25}\rangle$
           &$\left(\begin{array}{ccc} 0 & -\frac{7}{3\sqrt{5}}& \frac{2}{\sqrt{5}} \\ -\frac{7}{3\sqrt{5}} & \frac{16}{15}& -\frac{1}{5} \\ \frac{2}{\sqrt{5}} &-\frac{1}{5}& \frac{8}{5} \end{array}\right)$
           &$\left(\begin{array}{cccc} 0 & \frac{7}{3\sqrt{10}}& -\frac{16}{15}& -\frac{\sqrt{7}}{5\sqrt{2}}\\ \frac{7}{3\sqrt{10}} & 0& -\frac{7}{3\sqrt{10}} & -\frac{2}{\sqrt{35}} \\ -\frac{16}{15} & -\frac{7}{3\sqrt{10}}& 0& -\frac{1}{\sqrt{14}} \\-\frac{\sqrt{7}}{5\sqrt{2}}&-\frac{2}{\sqrt{35}} &-\frac{1}{\sqrt{14}}&\frac{4}{7}\end{array}\right)$
           &$\left(\begin{array}{cccc} 0 & \frac{2}{\sqrt{15}}& \frac{\sqrt{7}}{5\sqrt{3}}& -\frac{2\sqrt{14}}{5}\\ \frac{2}{\sqrt{15}} & 0& \frac{\sqrt{7}}{3\sqrt{5}} & -\frac{4\sqrt{2}}{\sqrt{105}} \\ \frac{\sqrt{7}}{5\sqrt{3}} & \frac{\sqrt{7}}{3\sqrt{5}}& -\frac{16}{21}& -\frac{\sqrt{2}}{7\sqrt{3}} \\-\frac{2\sqrt{14}}{5}&-\frac{4\sqrt{2}}{\sqrt{105}} &-\frac{\sqrt{2}}{7\sqrt{3}}&-\frac{4}{7}\end{array}\right)$\\
           \bottomrule[1pt]\bottomrule[1pt]
  \end{tabular}
\end{table*}

\end{document}